\documentclass[aps,twocolumn,prl,superscriptaddress]{revtex4}

\usepackage{graphicx}
\usepackage{amsmath}
\usepackage{amssymb}
\usepackage{units}
\usepackage{times}

\renewcommand{\vec}[1]{\boldsymbol{#1}}
\newcommand{\operator}[1]{\ensuremath{\hat{#1}}}

\begin{document}

%
%

\title{The Structural Influence on the Rashba-type Spin-Splitting in Surface Alloys}

\author{Isabella Gierz}
\email[Corresponding author; electronic address:\
]{i.gierz@fkf.mpg.de} \affiliation{Max-Planck-Institut f\"ur
Festk\"orperforschung, 70569 Stuttgart, Germany}
\author{Benjamin Stadtm\"uller}
\altaffiliation[Present address: ]{Institute of Bio- and
Nanosystems (IBN-3) and JARA-Fundamentals of Future Information
Technologies, Research Center J\"ulich, 52425 J\"ulich, Germany}
\affiliation{Max-Planck-Institut f\"ur Festk\"orperforschung,
70569 Stuttgart, Germany}\affiliation{Physikalisches Institut,
Universit\"at W\"urzburg, 97074 W\"urzburg, Germany}
\author{Johannes Vuorinen}
\affiliation{Tampere University of Technology, Deptartment of
Physics, 33101 Tampere, Finland}
\author{Matti Lindroos}
\affiliation{Tampere University of Technology, Deptartment of
Physics, 33101 Tampere, Finland}
\author{Fabian Meier}
\affiliation{Physik-Institut, Universit\"at Z\"urich, 8057
Z\"urich, Switzerland} \affiliation{Swiss Light Source, Paul
Scherrer Institute, 5232 Villigen, Switzerland}
\author{J.~Hugo Dil}
\affiliation{Physik-Institut, Universit\"at Z\"urich, 8057
Z\"urich, Switzerland} \affiliation{Swiss Light Source, Paul
Scherrer Institute, 5232 Villigen, Switzerland}
\author{Klaus Kern}
\affiliation{Max-Planck-Institut f\"ur Festk\"orperforschung,
70569 Stuttgart, Germany} \affiliation{IPMC, Ecole Polytechnique
F{\'e}d{\'e}rale de Lausanne, 1015 Lausanne, Switzerland}
\author{Christian R. Ast}
\affiliation{Max-Planck-Institut f\"ur Festk\"orperforschung,
70569 Stuttgart, Germany}

\date{\today}

\begin{abstract}
The Bi/Ag(111), Pb/Ag(111), and Sb/Ag(111) surface alloys exhibit
a two-dimensional band structure with a strongly enhanced
Rashba-type spin-splitting, which is in part attributed to the
structural asymmetry resulting from an outward relaxation of the
alloy atoms. In order to gain further insight into the
spin-splitting mechanism, we have experimentally determined the
outward relaxation of the alloy atoms in these surface alloys
using quantitative low-energy electron diffraction (LEED). The
structure plays an important role in the size of the
spin-splitting as it dictates the potential landscape, the
symmetry as well as the orbital character. Furthermore, we discuss
the band ordering of the Pb/Ag(111) surface alloy as well as the
reproducible formation of Sb/Ag(111) surface alloys with unfaulted
(face-centered cubic) and faulted (hexagonally close-packed)
toplayer stacking.
\end{abstract}

\maketitle

%
%

\section{INTRODUCTION}
The Rashba-Bychkov (RB) model has been remarkably successful in
describing two-dimensional (2D) electron systems with a structural
inversion asymmetry (SIA). It describes how an electric field
perpendicular to the 2D electron system lifts the spin-degeneracy
resulting in a characteristic band dispersion \cite{Rashba}.
Originally developed for semiconductor heterostructures, it has
also been successfully applied to surface states on various
surfaces even though the boundary conditions are different than
for the heterostructures \cite{LaShell}. Extending the RB model to
include an in-plane potential gradient coming from a lattice
potential with an in-plane inversion asymmetry, results in a
strong enhancement of the spin-splitting \cite{Premper}.

The spin-splitting mechanism relies on different contributions,
such as a strong spin-orbit interaction, a SIA as well as other
structural parameters (e.\ g., corrugation, relaxation, orbital
character) \cite{Petersen,Henk,Bihlmayer,Reinert,Dil,Huertas}. The
RB model can as such only serve as a qualitative model because it
only accounts for an effective electric field acting on a 2D free
electron gas, which combines all these different contributions in
one single parameter. The importance of detailed structural
considerations, which are completely neglected in the framework of
the RB model in the nearly free electron gas, becomes obvious when
comparing silver and antimony. Ag and Sb atoms are of about the
same size and exhibit a very similar atomic spin-orbit coupling.
The surface states on the (111) surfaces of these elements,
however, exhibit substantially different spin-splittings, the one
in Ag(111) being much smaller than in Sb(111)
\cite{Reinert,Sugawara,HugosScience}. Since surface potential
gradient and atomic spin-orbit coupling can be considered to be of
the same order of magnitude, the reason for this difference in
spin-splitting must be sought in the structure. The
face-centered-cubic structure of Ag features a smooth hexagonal
lattice at the (111) surface, whereas the rhombohedral structure
of Sb(111) with two atoms in the basis grows in a bilayer
configuration stacked along the [111] direction resulting in a
corrugated surface \cite{Liu}. Hence, structural considerations
have to be taken into account for a better understanding of the
Rashba-type spin-splitting in surface states. For systems with a
similar crystal structure like Bi and Sb the different size of the
spin-splitting can be attributed to the difference in atomic mass
\cite{Ast,Koroteev,Sugawara}. Further, even though the atomic
spin-orbit coupling is vanishingly small in graphene or carbon
nanotubes, it has been shown that the spin-orbit interaction can
be structurally enhanced by the local curvature in their structure
\cite{Huertas,Kuemmeth,Jeong}.

Recently a number of surface alloys (Bi/Ag(111), Pb/Ag(111), etc.)
have been identified with an extremely large spin-splitting in the
electronic structure at the surface
\cite{Pacile,Ast2,Ast3,Moreschini,Moreschini2}. Interestingly,
only a fraction of the atoms at the surface feature a sizeable
atomic spin-orbit coupling so that here as well the source of the
large spin-splitting must be sought in the structure. Motivated by
this and by theoretical considerations about the relaxation
dependence of the spin-splitting \cite{Bihlmayer} we present a
systematic study of the alloy atom relaxation in different Ag(111)
surface alloys using quantitative low-energy electron diffraction
(IV-LEED) measurements and calculations. We relate the dopant atom
relaxation to the size of the spin-splitting and address a number
of other unresolved issues in the literature. These include the
band assignment of the bands crossing the Fermi level in the
Pb/Ag(111) surface alloy, hcp (hexagonal close-packed) and fcc
(face-centered cubic) toplayer stacking in the Sb/Ag(111) surface
alloy as well as the Bi atom relaxation in the Ag(111) and Cu(111)
surface alloys.

\section{EXPERIMENTS AND CALCULATIONS}
All experiments were performed in ultra high vacuum (UHV) with a
base pressure of $1\times10^{-10}$\,mbar at 77\,K. The angular
resolved photoemission spectroscopy (ARPES) measurements were done
with a SPECS HSA3500 hemispherical analyzer with an energy
resolution of 10\,meV and monochromatized HeI radiation. The
IV-LEED measurements where done using an ErLEED 1000-A. During
IV-LEED measurements magnetic stray fields were compensated using
Helmholtz coils to assure perpendicular incidence of the electrons
on the sample for all kinetic energies. The Ag(111) substrate was
cleaned using several sputtering/annealing cycles. Cleanliness of
the substrate was checked with X-ray photoemission spectroscopy
(XPS) and the different surface states were controlled with ARPES.
One third of a monolayer (ML) of Bi, Pb, or Sb (henceforth
referred to as alloy atoms) was deposited using a commercial
electron beam evaporator. The substrate temperature during
deposition was 420$^{\circ}$C, 350$^{\circ}$C, and 250$^{\circ}$C
for Bi, Pb and Sb, respectively.

We performed IV-LEED calculations using the experimental data for
Bi/Ag(111), Pb/Ag(111) and two different measurements for
Sb/Ag(111). For each of the four experimental data sets both the
unfaulted (fcc) substitutional and the faulted (hcp)
substitutional structures were calculated. For all structures the
decision between fcc and hcp toplayer stacking was very clear as
the Pendry $R_P$-factors for the rejected structures were in all
cases more than double to those of the correct structures.

During the calculations surface geometry, Debye temperatures and
real part of the inner potential were optimized. In every set the
geometry of the four highest layers was calculated as a
$(\sqrt{3}\times\sqrt{3})$ unit cell. The symmetry restricts the
degrees of freedom to the component normal to the surface and also
forbids buckling for most layers. Three Debye temperatures $T_D$
were fitted for each measurement, the top layer Ag atoms, top
layer Pb/Bi/Sb atoms and of all the other Ag atoms. The real part
of the inner potential was optimized and found to be approximately
6\,eV for all calculations. The imaginary part of the potential
was kept at a fixed value of 4.5\,eV for all calculations.

The LEED calculations were performed using the Barbieri/Van Hove
SATLEED package \cite{satleed}. The necessary fourteen phase
shifts were determined with the Barbieri/Van Hove phase shift
package \cite{Barbieri}. Temperature effects were calculated
within the SATLEED code by multiplying each atoms scattering
amplitude by a Debye-Waller factor. Pendry $R_P$-factors
\cite{Pendry1980} were used to measure the level of agreement
between measured and calculated IV-LEED spectra and statistical
errors in analysis were estimated with Pendry RR-factors
\cite{Pendry1980}.

\section{RESULTS AND DISCUSSION}

\begin{figure}
  \includegraphics[width = 1\columnwidth]{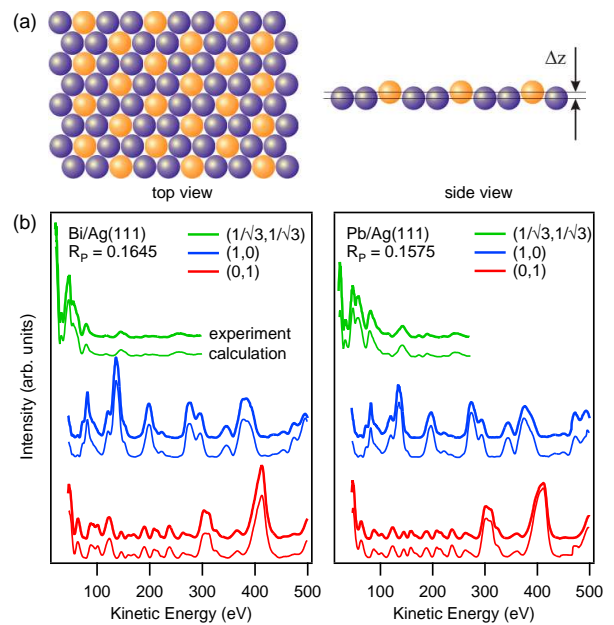}
  \caption{(color online) Panel (a) shows the top and side views for the
  ($\sqrt3\times\sqrt3$)R30$^{\circ}$ structure formed by the surface
  alloys on Ag(111). Ag atoms are shown in red, alloy atoms are
  shown in blue. The outward relaxation of the alloy atoms is defined as the
  vertical distance between alloy and Ag atoms in the topmost layer.
  In Panel (b) IV-LEED data (thick lines) for the ($\sqrt3\times\sqrt3$)R30$^{\circ}$
  phase of Bi (left) and Pb (right) on Ag(111) are displayed
  together with the corresponding calculated IV-LEED spectra (thin lines).}
  \label{Fig.1}
\end{figure}

Although (Bi, Pb, Sb) and Ag are immiscible in the bulk, they form
a long-range ordered surface alloy where every third Ag atom in
the topmost layer is replaced by an alloy atom. The resulting
($\sqrt{3}\times\sqrt{3}$)R30$^{\circ}$ structure (with respect to
the pristine Ag(111) surface) is displayed in Fig. \ref{Fig.1} a).
The alloy atoms relax out of the plane of the Ag layer by an
amount $\Delta z$ as shown in the side view in Fig.\ \ref{Fig.1}
a) \cite{Oppo,Dalmas,Pacile,Ast2,Bihlmayer}.

\begin{figure}
  \includegraphics[width = \columnwidth]{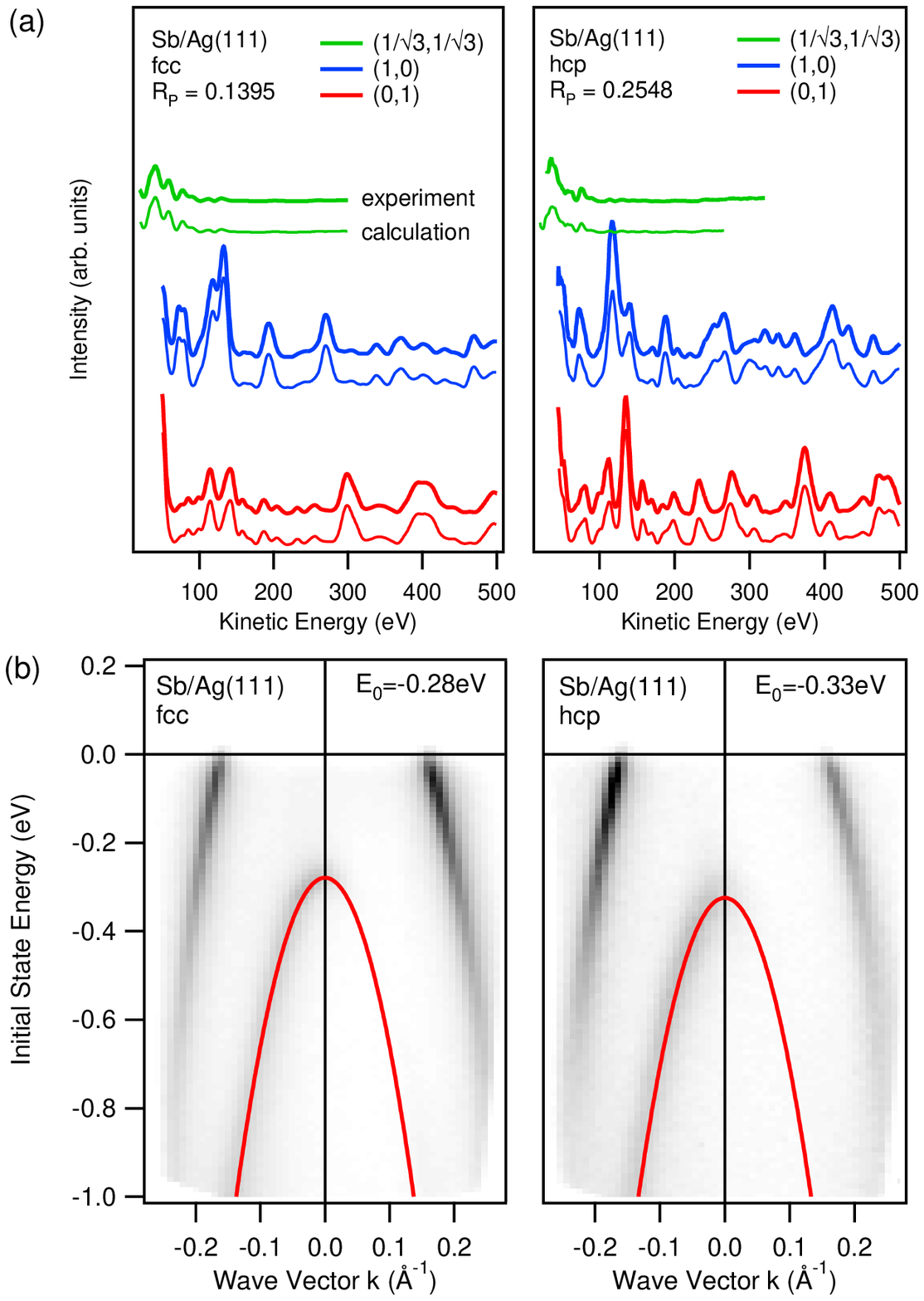}
  \caption{(color online) Comparison of IV-LEED (a) and ARPES (b)
  data for fcc (left panels) and hcp (right panels) toplayer stacking
  for the Sb/Ag(111) surface alloys. The red lines in b) are fits to
  the experimental data.}
  \label{Fig.2}
\end{figure}

Fig.\ \ref{Fig.1} b) shows the integrated intensity of the (0,1),
(1,0), and (1/$\sqrt{3}$,1/$\sqrt{3}$) spots as a function of
electron energy for the Bi/Ag(111) and Pb/Ag(111) surface alloys.
The data were averaged over three (six) equivalent spots and
smoothed. The IV-LEED spectra for both Bi/Ag(111) and Pb/Ag(111)
surface alloys differ only in detail indicating that they form the
same ($\sqrt{3}\times\sqrt{3}$)R30$^{\circ}$ structure. For the
Sb/Ag(111) surface alloy two different phases are known from the
literature, which differ in the toplayer stacking
\cite{Woodruff,Soares,Quinn}. One phase grows in regular
face-centered cubic (fcc) stacking while the other phase grows in
hexagonal close-packed (hcp) stacking. The IV-LEED spectra for the
two Sb/Ag(111) surface alloys, which have been treated in analogy
to the spectra in Fig.\ \ref{Fig.1} b), are shown in Fig.\
\ref{Fig.2} a). The thick and thin lines in the graphs correspond
to the experimental and calculated spectra, respectively. It can
be seen that for all surface alloys the calculated spectra fit
very well to the experimental data resulting in low $R_P$-factors.
Only the $R_P$-factor for the hcp-phase of the Sb/Ag(111) is
somewhat higher, which we attribute to some fcc-domains being
present at the surface.

\begin{table*}
\caption{Geometrical parameters of the different surface alloys on
Ag(111) and Cu(111) substrates. The outward relaxation $\Delta z$
is the distance between the alloy atom and the plane of the
surface layer. The distances $\mathbf{d_{12}}$, $\mathbf{d_{23}}$,
and $\mathbf{d_{34}}$ are the distances between the first (=
surface) and second layer, the second and third layer, as well as
the third and fourth layer, respectively. The bulk interlayer
distances are 2.36\,\AA\ for Ag(111) and 2.09\,\AA\ for Cu(111).}
\label{Tab:IVLEEDresults}
\begin{tabular*}{\textwidth}{c@{\extracolsep\fill}clccccc}
\hline\hline
&\textbf{geometry}&$\mathbf{\Delta z}$ \textbf{(\AA)}&$\mathbf{d_{12}}$ \textbf{(\AA)}&$\mathbf{d_{23}}$ \textbf{(\AA)}&$\mathbf{d_{34}}$ \textbf{(\AA)}&$\mathbf{R_P}$ & Ref.\\
\textbf{Bi/Ag(111)}&substitutional&$0.65\pm0.10$&$2.32\pm0.02$&$2.33\pm0.03$&$2.34\pm0.04$&$0.1645$& this work\\
       & & 0.35 (theory)  &   &   &   &  & \cite{Ast}\\
       & & 0.85 (theory)  &   &   &   &  & \cite{Bihlmayer}\\
\textbf{Pb/Ag(111)}&substitutional&$0.46\pm0.06$&$2.35\pm0.02$&$2.33\pm0.03$&$2.34\pm0.04$&$0.1575$ & this work\\
       & & 0.24 (XRD) &   &   &  &  & \cite{Dalmas} \\
       & & 0.8 (STM) &   &   &  &   & \cite{Dalmas} \\
       & & 0.68 (theory) &   &   &   &  & \cite{Dalmas} \\
       & & 0.42 (theory) &   &   &   &  & \cite{Ast2} \\
       & & 0.97 (theory) &   &   &   &  & \cite{Bihlmayer} \\
\textbf{Sb/Ag(111)}&hcp substitutional&$0.11\pm0.05$&$2.43\pm0.05$&$2.34\pm0.05$&$2.35\pm0.06$&$0.2548$ & this work\\
       & & $0.03\pm0.07$ (XRD) & $2.50\pm0.03$  &  &  &  & \cite{deVries}\\
       & & 0.02 (theory) &  &  &  &  & \cite{Woodruff}\\
       & & 0.07 (LEED) &  &  &  &  & \cite{Soares} \\
\textbf{Sb/Ag(111)}&substitutional&$0.10\pm0.02$&$2.44\pm0.02$&$2.33\pm0.02$&$2.33\pm0.03$&$0.1395$ & this work\\
       & & 0.24 (theory) &  &  &  &  & \cite{Moreschini}\\
\textbf{Bi/Cu(111)}&substitutional&$1.02\pm0.02$ (XRD)&$2.12\pm0.01$&$2.10\pm0.01$ & & & \cite{Kaminski}\\
       & & $1.06$ (theory) &  &  &  &  & \cite{Bentmann}\\
\textbf{Sb/Cu(111)}&hcp substitutional&$0.47\pm0.16$ (MEIS)&$2.05\pm0.09$& & & & \cite{Bailey}\\
       & & $0.6\pm0.03$ (XRD) & $1.98\pm0.02$  &  &  &  & \cite{deVries}\\

\hline\hline
\end{tabular*}
\end{table*}

The structural parameters resulting from the calculated IV-LEED
spectra are summarized in Table \ref{Tab:IVLEEDresults} along with
some values from the literature for comparison. The outward
relaxation $\Delta z$, which is the distance between the alloy
atom and the plane of the first layer of substrate atoms (see
Fig.\ \ref{Fig.1}b), is given for each of the different alloy
atoms. In addition, the interlayer distances for the first four
layers are summarized. After the fourth layer no significant
deviation from the bulk value of 2.36\,\AA\ is expected. The
in-plane lattice constants were held fixed. While for the
Pb/Ag(111) and Bi/Ag(111) surface alloys the interlayer distances
hardly differ from the bulk, it should be noted that in the
Sb/Ag(111) surface alloy the distance between the first and the
second silver layer is increased by about 0.1\,\AA. Furthermore,
non-structural parameters that accompany the IV-LEED calculations
such as the Debye temperatures as well as the real part of the
inner potential are given in Table \ref{Tab:IVLEEDnonstruct}. As
the IV-LEED measurements were done at 77\,K, which is comparable
to the low Debye temperatures, no complications from a too high
Debye-Waller factor were anticipated in the calculations.

In the following, we relate the structural findings to the size of
the Rashba-type spin-splitting. The RB model is described by the
Hamiltonian:
\begin{equation}
  \operator{H}_{\mathrm{SO}}
  =
  \alpha_{\mathrm{R}}
  \vec{\sigma}
  \cdot
  (\vec{k}_{\parallel} \times \vec{e}_{z})
  \label{eq:Ham}
\end{equation}
where $\alpha_{\mathrm{R}}$ is the Rashba parameter. The resulting
energy dispersion in the nearly free electron model is
$E(\vec{k}_{||})=\frac{\hbar^2}{2m^*}(k_{||}\pm k_0)^2 + E_0$
where $m^*$ is the effective mass, $k_0=m^*\alpha_R/\hbar^2$ is
the offset by which the parabola is shifted away from the high
symmetry point, and $E_0$ is an offset in energy. The Rashba
energy $E_R=\hbar^2k_0^2/2m^*$ is the energy difference between
the band extremum and the crossing point at the high symmetry
point. The three parameters $k_0$, $\alpha_R$, and $E_R$ quantify
the strength of the Rashba-type spin-splitting and serve well to
compare the different spin-split bands in the surface alloys. An
overview of the typical parameters in these systems is given in
Table \ref{table1}.

\begin{table*}
\caption{Non structural parameters that have been used in the
IV-LEED calculation. The Debye temperatures $T_D$ for the
different alloy atoms (Pb/Bi/Sb), the surface layer silver atoms
(Ag$_1$), and the silver atoms in the subsequent layers
(Ag$_{2\rightarrow}$) are given along with the real part of the
inner potential $V_R$. }\label{Tab:IVLEEDnonstruct}
\begin{tabular*}{\textwidth}{c@{\extracolsep\fill}cccc}
\hline\hline
\strut\hfill\strut&$\mathbf{T_D(Pb/Bi/Sb) (K)}$&$\mathbf{T_D(Ag_1) (K)}$&$\mathbf{T_D(Ag_{2\rightarrow})(K)}$&$V_R (\text{eV})$\\
\textbf{Pb/Ag(111)}&90&150&180&6.3\\
\textbf{Bi/Ag(111)}&65&160&210&5.9\\
\textbf{Sb/Ag(111) hcp subst.}&150&110&230&5.7\\
\textbf{Sb/Ag(111) subst.}&115&130&200&6.2\\
\hline\hline
\end{tabular*}
\end{table*}

\subsection{Bi/Ag(111)}

With $\Delta z=0.65\pm0.10$\,\AA, the Bi/Ag(111) surface alloy
shows the largest outward relaxation in the surface alloys
considered here on a Ag(111) substrate. So far only two
theoretical values have been reported for the relaxation, one of
which is smaller while the other is larger than the experimental
value (see Table \ref{Tab:IVLEEDresults}). The difference between
the theoretical values is most likely related to different methods
for relaxing the structure. Nevertheless, the corresponding band
structure calculations both show good agreement with the
experimental data \cite{Bihlmayer,Ast2}.

\begin{table}
\caption{Characteristic parameters for the spin-split states in
the different surface alloys on the Ag(111) and Cu(111) surfaces.
The different parameters, momentum offset $k_0$, Rashba energy
$E_R$, and Rashba constant $\alpha_R$ are defined in the text.}
\begin{tabular*}{\columnwidth}{c@{\extracolsep\fill}cccc}
 \hline\hline
 System & $k_0$ ({\AA}$^{-1}$) & $E_R$ (meV) & $\alpha_R$ (eV{\AA}) & Ref. \\
 Sb/Ag(111) & 0.005  & 5.7  & 0.76 & \cite{Meier3,Moreschini}\\
 Pb/Ag(111) & 0.03 & 23 & 1.52  & \cite{Ast3} (exp) \\
            & 0.04 & 22 & 1.05  & \cite{Ast3} (theor) \\
            & 0.11 &    &       & \cite{Bihlmayer} \\
 Bi/Ag(111) & 0.13 & 200 & 3.05 & \cite{Ast2} \\
            & 0.13 &     &      & \cite{Bihlmayer}\\
 Sb/Cu(111) & 0.005 & 3 & 0.19 & \cite{SbCuSpinSplitting}\\
 Bi/Cu(111) & 0.03 & 15 & 1.0 & \cite{Moreschini2}\\
            & 0.096 &   &      & \cite{Mirhosseini}\\
            & 0.032 & 13 & 0.82 & \cite{Bentmann} (exp)\\
            & 0.028   &  9 & 0.62 & \cite{Bentmann} (theor)\\
 Au(111)    & 0.012 & 2.1 & 0.33 & \cite{Cercellier}\\
 Ag(111)    & 0.0007  & 0.005 & 0.013 & \cite{Reinert} \\
 Cu(111)    & 0 & 0 & 0 & \cite{Cu}  \\
 \hline\hline
\end{tabular*}
\label{table1}
\end{table}

Calculations have shown that an increased $p_{xy}$-character in
the fully occupied band is responsible for the enhanced
spin-splitting characterized by a momentum offset of
$k_0=0.12$\,\AA$^{-1}$ in Bi/Ag(111) \cite{Bihlmayer}. It was
shown that starting from a hypothetical flat Bi/Ag(111) surface
alloy the $s:p_z$-ratio changes in favor of the $p_z$-character
with an admixture of $p_{xy}$-character upon outward relaxation of
the Bi atoms resulting in a stronger spin-splitting. Comparing the
Bi/Ag(111) surface alloy to the Bi/Cu(111) surface alloy, we find
that the structure is qualitatively the same for both surface
alloys. However, for the Bi/Cu(111) surface alloy outward
relaxations of more than one \AA ngstr\"om have been reported (see
Table \ref{Tab:IVLEEDresults}). The difference in outward
relaxation can be attributed to the smaller lattice constant of
the Cu substrate. A comparison of the electronic structures shows
that the spin-splitting $k_0=0.03$\,\AA$^{-1}$ for Bi/Cu(111) is
much smaller than for Bi/Ag(111). Furthermore, the surface alloy
bands in Bi/Cu(111) are shifted into the unoccupied states, which
can be related to a different charge transfer to the surface state
from the Cu bulk as compared to the Ag bulk.

\subsection{Pb/Ag(111)}

For the Pb/Ag(111) surface alloy an outward relaxation of $\Delta
z=0.46\pm0.06$\,\AA\ has been found. This value lies in between
previously reported values excluding the value obtained by
scanning tunneling microscopy (STM). STM is generally not very
suitable for obtaining structural parameters perpendicular to the
surface \cite{Dalmas}. Also, in contrast to what has been observed
before \cite{Dalmas}, our IV-LEED calculations show that the two
lattice sites of the Ag atoms in the
$\sqrt{3}\times\sqrt{3}R30^{\circ}$ unit cell are equivalent.

The size of the spin-splitting for the Pb/Ag(111) surface alloy is
still under debate due to different interpretations of the
measured ARPES spectra \cite{Ast4,Bihlmayer,Hirahara,Meier}.
Depending on the band dispersion of the $sp_z$-band and the
$p_{xy}$-band in the unoccupied states, the occupied states may be
assigned to different bands \cite{Bihlmayer,Ast2}. It has been
shown that the position of the $sp_z$-band is particularly
sensitive to the outward relaxation of the Pb atoms
\cite{Bihlmayer}. With a calculated outward relaxation of
0.97\,{\AA}, the $sp_z$-band with $k_0$=0.11\,\AA$^{-1}$ and the
$p_{xy}$-band cross without hybridizing (Scenario I). By reducing
the relaxation of the Pb atoms to 0.67\,\AA, however, the two sets
of bands avoid each other resulting in a lower apparent
spin-splitting (Scenario II) \cite{Bihlmayer}. Other calculations
found a relaxation of 0.42\,\AA\ leading to a spin-splitting of
0.04\,\AA$^{-1}$ in good agreement with experiment
\cite{Ast3,Meier2}. Here, no band crossing between the $sp_z$-band
and the $p_{xy}$-band has been observed. As the experimental
outward relaxation found for Pb/Ag(111) is 0.46\,\AA, we are
inclined to favor the second scenario. This interpretation is
further supported by recent spin-resolved ARPES measurements
\cite{Meier}.

\begin{figure*}
  \includegraphics[width = 1\textwidth]{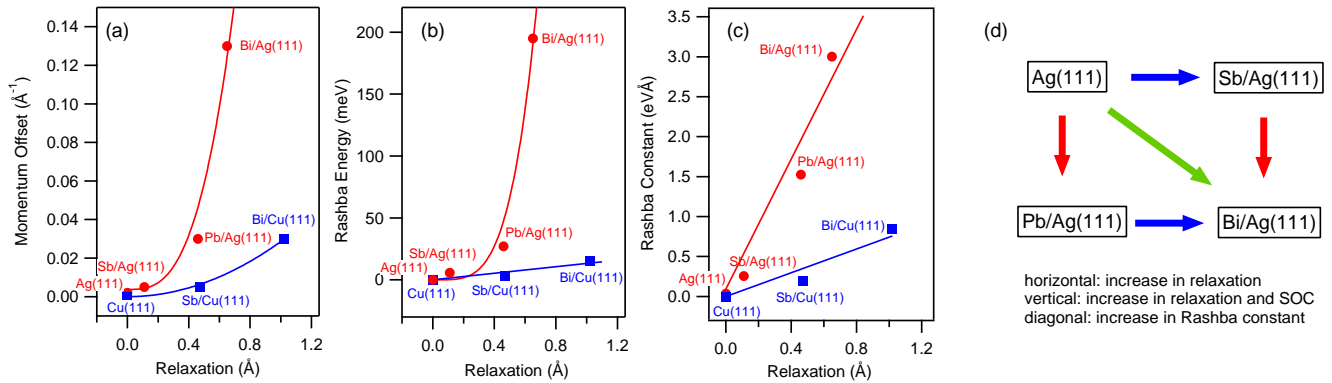}
  \caption{(color online) Characteristic experimental parameters for a Rashba system, such as
  momentum offset (a), Rashba Energy (b), and Rashba constant are
  shown as a function of the outward relaxation $\Delta z$. The
  lines are drawn as a guide to the eye. In (d) a schematic of the
  evolution of the outward relaxation and the atomic spin-orbit coupling
  is drawn for the different surface alloys.}
  \label{Fig.4}
\end{figure*}

\subsection{Sb/Ag(111)}

The Sb/Ag(111) surface alloy can be formed with either fcc or hcp
toplayer stacking \cite{Woodruff,Quinn}. However, no reproducible
way of creating these two phases has been reported so far. The
formation of the hcp toplayer stacking has been attributed to the
presence of subsurface stacking faults from previous preparations
caused by Sb atom diffusion into the bulk \cite{Woodruff}. We have
found that it is possible to reproducibly create the two phases
separately regardless of the sample preparation history. In the
electron beam evaporator used for depositing the Sb atoms, the
atom beam is partially ionized by electrons from a filament. A
positive voltage at the crucible accelerates the Sb ions towards
the grounded sample. The higher the voltage at the crucible, the
higher the kinetic energy of the Sb ions and therefore the
stronger their impact at the Ag(111) surface. For voltages below
$+370$\,V at the Sb crucible as well as for thermal Sb atoms
evaporated from a Knudsen cell, we found that the phase with
fcc-stacking is formed. For higher voltages the phase with
hcp-stacking is formed. It is conceivable that the higher ion
energy results in ion implantation into the Ag substrate inducing
the subsurface stacking faults that favor the phase with
hcp-stacking.

The Sb outward relaxation for the fcc- and hcp-phase is
$0.1\pm0.02$\,\AA\ and $0.11\pm0.05$\,\AA, respectively. These
values are similar to what has been previously found for the
hcp-phase using IV-LEED. In addition, a slight outward relaxation
between 0.07\,\AA\ and 0.08\,\AA\ of the surface Ag layer has been
observed for both phases. The Ag layer relaxation has also been
found by x-ray diffraction (XRD). There the Ag layer relaxation is
larger (0.14\,\AA), whereas the outward relaxation of the Sb atom
is smaller (0.03\,\AA) \cite{deVries}. A similar effect has been
observed for the Sb/Cu(111) surface alloy. There, however, the Cu
layer relaxation reduces the distance between the surface layer
and the substrate \cite{deVries,Bailey}.

The toplayer stacking also slightly affects the band structure.
This can be seen in Fig.\ \ref{Fig.2}(b), where the experimental
band structure of the two phases for the Sb/Ag(111) surface alloy
in the vicinity of the $\overline{\Gamma}$-point is shown. We
observe a small shift of 50\,meV between the $sp_z$-band of the
two phases. However, due to the comparatively large error bars we
cannot relate this to the outward relaxation. We rather attribute
this shift to possible disorder at the surface. This
interpretation is also supported by the larger $R_P$-factor for
the hcp-phase.

The spin-splitting of the sp$_z$-band is too small to be detected
by conventional ARPES. However, spin-resolved photoemission
experiments, which are much more sensitive to a small
spin-splitting due to the ``spin-label'' of the bands
\cite{Meier}, show that the $sp_z$-band of the hcp-phase in the
Sb/Ag(111) surface alloy is spin-split by 0.005\,\AA$^{-1}$
\cite{Meier3}. Unfortunately, no spin-splitting has been reported
so far for the fcc-phase.

\subsection{Spin-splitting vs. Relaxation}

The large spin-splitting in the different surface alloys cannot be
accounted for by the spin-orbit interaction in the heavy elements
alone. This becomes obvious when comparing them to other materials
with a sizeable spin-splitting, such as Au(111) or Bi(111). The
spin-splitting of the surface state on the pristine Ag(111)
substrate is experimentally undetectable with current experimental
techniques. The spin-splitting in the Bi/Ag(111) surface alloy is
much larger than what has been observed for the Bi(111) surface
state, even though only a fraction of the atoms in the surface
alloy are Bi atoms. Thus, the simple statement that a strong
spin-orbit coupling leads to a large spin-splitting does not hold.

In order to resolve this issue the structural details of the
different materials have to be considered. As the structure
dictates the potential landscape, the orbital overlap as well as
the orbital hybridization, it has direct influence on the
asymmetry of the wave functions and the corresponding
spin-splitting in the electronic structure. In Fig.\
\ref{Fig.4}a--c the characteristic parameters for the Rashba-type
spin-splitting in the different surface alloys --- momentum offset
$k_0$, Rashba energy $E_R$, and Rashba constant $\alpha_R$ --- are
plotted as a function of the outward relaxation. The surface
alloys on the Cu(111) and the Ag(111) substrates are drawn in blue
squares and red circles, respectively. The solid lines are drawn
to guide the eye.

Unfortunately, with experimentally available systems it is
difficult to change only one parameter, while leaving everything
else constant. Nevertheless, the available data allows us to
define some trends. A schematic of the different parameter changes
is shown in Fig.\ \ref{Fig.4}d for the four different systems that
are known on a Ag(111) substrate. The bare Ag(111) surface state
has been included as a reference point with no outward relaxation.
The other three surface alloys are arranged such that the
parameter changes become apparent. Along the horizontal direction,
from left to right the outward relaxation changes, but the mass of
the elements (Ag $\rightarrow$ Sb; Pb $\rightarrow$ Bi) remains
almost the same. Along the vertical direction, the outward
relaxation changes along with the atomic spin-orbit coupling as
the atomic mass changes by a factor of almost two (Ag
$\rightarrow$ Pb; Sb $\rightarrow$ Bi).

With the schematic in Fig.\ \ref{Fig.4}d in mind, the general
trend that an increased spin-orbit coupling and an increased
outward relaxation leads to an increased spin-splitting seems to
be evident, when considering the Ag(111) and the Cu(111)
substrates separately. If we consider that the atomic spin-orbit
coupling constant for $p$-electrons in a hydrogen-like atom is
proportional to $Z^4/n^3$ ($Z$: atomic number, $n$: principal
quantum number), the increase in the atomic spin-orbit coupling
constants of Sb, Pb, and Bi alone cannot account for the increase
in the respective Rashba constants. Therefore, we conclude that
the structure of the surface alloy, i.\ e.\ the outward
relaxation, plays a key role in the strength of the
spin-splitting.

Comparing the surface alloys for the Ag(111) and Cu(111)
substrates, the situation is not so straightforward anymore. The
Bi/Cu(111) surface alloy shows a much larger outward relaxation,
but a much smaller spin-splitting than the Bi/Ag(111) surface
alloy. It is conceivable that the substrate itself has an
influence on the size of the spin-splitting. Calculations for the
$sp_z$-states have shown that the substrate atoms within the
surface alloy layer carry a significant spectral weight, which has
lead to the conclusion that the atomic spin-orbit parameter of the
substrate contributes to the spin-splitting in the corresponding
surface alloys \cite{Moreschini2}. However, in another comparison
of the Bi/Ag(111) and the Bi/Cu(111) surface alloy, the spin-orbit
coupling of the substrate has been found to play a negligible role
in the spin-splitting of the surface alloy. The conclusion here
was that structural effects (i.\ e.\ the outward relaxation of the
alloy atom) changing the orbital composition play a dominant role
\cite{Bentmann}. The difference between Bi/Ag(111) and Bi/Cu(111)
could simply originate from the different lattice constants of the
substrate. In this regard, a simple tight-binding calculation
shows that the Rashba constant is proportional to the lattice
constant \cite{propto}. Another possible explanation is that the
spin-splitting reaches a maximum with further outward relaxation
leading to smaller spin-splitting. In addition, the distance
$d_{12}$ between the surface layer and the substrate is compressed
for Bi/Ag(111) and expanded for Bi/Cu(111) with respect to the
bulk interlayer spacing. Whether this plays a role in the
spin-splitting for the surface alloys so far is unknown.

The results discussed here show that the outward relaxation plays
a key role in the size of the spin-splitting in the surface alloys
with the general trend that a larger outward relaxation leads to a
larger spin-splitting. However, many open questions remain, such
as a better understanding of the role of the substrate, the
influence of avoided crossings of the bands and the resulting
apparent spin-splittings as well as the question whether there is
a maximum in outward relaxation after which the spin-splitting
decreases again.

\section{Conclusion}

We have experimentally determined the outward relaxation of the
alloy atoms for three different surface alloys (Bi/Ag(111),
Pb/Ag(111) and Sb/Ag(111)) employing quantitative LEED
measurements and calculations. The outward relaxation for
Pb/Ag(111) is 0.46\,\AA, which leads us to favor the second
scenario with the smaller spin-splitting. In addition, we found
that the Sb/Ag(111) surface alloy can be grown reproducibly in
fcc- and hcp-toplayer stacking. Furthermore, we have related the
outward relaxation to the strongly enhanced spin-splitting in the
Ag(111) surface alloys comparing them also to surface alloys found
on Cu(111). We find that the outward relaxation plays an extremely
important role in the size of the spin-splitting, because the
ratio of the spin-orbit coupling strengths alone does not account
for the ratio of the Rashba constants in two different surface
alloys. Looking at each substrate individually a clear trend that
a large outward relaxation leads to a large spin-splitting is
evident. Deviations from this trend can be observed when comparing
the surface alloys on two different substrates, e.\ g.\ Bi/Ag(111)
and Bi/Cu(111). This could be explained by the different orbital
composition in the surface alloy band structure. The role of the
substrate has not been completely solved yet.

We conclude that the structure plays an important role in the
spin-splitting as it defines the potential landscape and has a
profound influence on the orbital overlap and the band dispersion.
This has also been found for graphene as well as carbon nanotubes
where the atomic spin-orbit interaction is clearly small.
Nevertheless, a straightforward, intuitive model for a better
understanding of the Rashba-type spin-splitting at surfaces would
be desirable --- if it exists.

\section{ACKNOWLEDGMENTS}
We would like to thank U.\ Starke for help with the IV-LEED
measurements as well as H.\ M.\ Benia for stimulating discussions.
C.\ R.\ A.\ acknowledges funding from the Emmy-Noether-Program of
the Deutsche Forschungsgemeinschaft (DFG).


\begin{thebibliography}{12}
 \bibitem{Rashba} Y. A. Bychkov and E. I. Rashba, Sov. Phys. JETP Lett. {\bf 39}, 78 (1984)
 \bibitem{LaShell} S. LaShell, B. A. McDougall and E. Jensen, Phys. Rev. Lett. {\bf 77}, 3419 (1996)
 \bibitem{Premper} J. Premper, M. Trautmann, J. Henk and P. Bruno, Phys. Rev. B {\bf 76}, 073310 (2007)
 \bibitem{Petersen} L. Petersen and P. Hedeg{\aa}rd, Surf. Science {\bf 459} 49 (2000)
 \bibitem{Henk} J. Henk, A. Ernst, P. Bruno, Phys. Rev. B {\bf 68}, 165416 (2003)
 \bibitem{Bihlmayer} G. Bihlmayer, S. Bl\"ugel and E. V. Chulkov {\it Phys. Rev. B} {\bf 75} 195414 (2007)
 \bibitem{Reinert} F. Reinert, J. Phys.: Cond. Mat. {\bf 15}, S693 (2003)
 \bibitem{Dil} J.~H. Dil, J. Phys.: Cond. Mat. {\bf 21}, 403001 (2009)
 \bibitem{Huertas} D. Huertas-Hernando, F. Guinea, A. Brataas, Phys. Rev. B {\bf 74}, 155426 (2006)
 \bibitem{Sugawara} K. Sugawara, T. Sato, S. Souma, T. Takahashi, M. Arai, and T. Sasaki, Phys. Rev. Lett. {\bf 96}, 046411 (2006)
 \bibitem{HugosScience} D. Hsieh, Y. Xia, L. Wray, D. Qian, A. Pal, J.~H. Dil, J. Osterwalder, F. Meier, G. Bihlmayer, C.~L. Kane, Y.~S. Hor, R.~J. Cava, M.~Z. Hasan, Science {\bf 323}, 919 (2009)
 \bibitem{Liu} Y. Liu and R.~E. Allen, Phys. Rev. B {\bf 52}, 1566 (1995) and references therein
 \bibitem{Ast} C.~R. Ast, H. H{\"o}chst, Phys. Rev. Lett. {\bf 87}, 177602 (2001)
 \bibitem{Koroteev} Y.~M. Koroteev, G. Bihlmayer, J.~E. Gayone, E.~V. Chulkov, S. Bl\"ugel, P.~M. Echenique, P. Hofmann, Phys. Rev. Lett. {\bf 93}, 046403 (2004)
 \bibitem{Kuemmeth} F. Kuemmeth, S. Ilani, D.~C. Ralph, and P.~L. McEuen, Nature {\bf 452}, 448 (2008)
 \bibitem{Jeong} J.-S. Jeong, H.-W. Lee, Phys. Rev. B {\bf 80}, 075409 (2009)
 \bibitem{Pacile} D. Pacil\'e, C. R. Ast, M. Papagno, C. Da Silva, L. Moreschini, M. Falub, A. P. Seitsonen and M. Grioni {\it Phys. Rev. B} {\bf 73} 245429 (2006)
 \bibitem{Ast2} C. R. Ast, J. Henk, A. Ernst, L. Moreschini, M. C. Falub, D. Pacil\'e, P. Bruno, K. Kern and M. Grioni {Phys. Rev. Lett.} {\bf 98} 186807 (2007)
 \bibitem{Ast3} C. R. Ast, D. Pacil\'e, L. Moreschini, M.~C. Falub, M. Papagno, K. Kern, M. Grioni, J. Henk, A. Ernst, S. Ostanin and P. Bruno, {\it Phys. Rev. B} {\bf 77}, 081407 (R) (2008)
 \bibitem{Moreschini} L. Moreschini, A. Bendounan, I. Gierz, C. R. Ast, H. Mirhosseini, H. H\"ochst, K. Kern, J. Henk, A. Ernst, S. Ostanin, F. Reinert and M. Grioni, {\it Phys. Rev. B} {\bf 79} 075424 (2009)
 \bibitem{Moreschini2} L. Moreschini, A. Bendounan, H. Bentmann, M. Assig, K. Kern, F. Reinert, J. Henk, C. R. Ast and M. Grioni {\it Phys. Rev. B} {\bf 80} 035438 (2009)
 \bibitem{satleed} A. Barbieri and M.~A. Van Hove, Symmetrized Automated Tensor LEED package, available from M.~A. Van Hove
 \bibitem{Barbieri} A. Barbieri and M.~A. Van Hove, private communication
 \bibitem{Pendry1980} J.~B.Pendry, J. Phys. C: Solid St. Phys. {\bf 13}, 937 (1980)
 \bibitem{Dalmas} J. Dalmas, H. Oughaddou, C. L\'eandri, J.-M. Gay, G. Le Lay, G. Tr\'eglia, B. Aufray, O. Bunk and R. L. Johnson {\it Phys. Rev. B} {\bf 72} 155424 (2005)
 \bibitem{Oppo} S. Oppo, V. Fiorentini, and M. Scheffler, Phys. Rev. Lett. {\bf 71}, 2437 (1993)
 \bibitem{Woodruff} D. P. Woodruff and J. Robinson {\it J. Phys.: Condens. Matter} {\bf 12} 7699 (2000)
 \bibitem{Soares} E. A. Soares, C. Bittencourt, V. B. Nascimento, V. E. de Carvalho, C. M. C. de Castilho, C. F. McConville, A. V. de Carvalho, D. P. Woodruff {\it Phys. Rev. B} {\bf 61} 13983 (2000)
 \bibitem{Quinn} P. D. Quinn, D. Brown, D. P. Woodruff, P. Bailey and T.~C.~Q. Noakes {\it Surf. Science} {\bf 511} 43 (2002)
 \bibitem{Hirahara} T. Hirahara, T. Komorida, A. Sato, G. Bihlmayer, E. V. Chulkov, K. He, I. Matsuda and S. Hasegawa {\it Phys. Rev. B} {\bf 78} 035408 (2008)
 \bibitem{Ast4} C. R. Ast, G. Wittich, P. Wahl, R. Vogelgesang, D. Pacil\'e, M.~C. Falub, L. Moreschini, M. Papagno, M. Grioni, and K. Kern, Phys. Rev. B {\bf 75}, 201401 (R) (2007)
 \bibitem{Meier} F. Meier, H. Dil, J. Lobo-Checa, L. Patthey and J. Osterwalder {\it Phys. Rev. B} {\bf 77} 165431 (2008)
 \bibitem{Meier2} F. Meier, V. Petrov, S. Guerrero, C. Mudry, L. Patthey, J. Osterwalder, J.~H. Dil, Phys. Rev. B {\bf 79}, 241408 (R) (2009)
 \bibitem{deVries} S. A. de Vries, W. J. Huisman, P. Goedtkindt, M. J. Zwanenburg, S. L. Bennett, I. K. Robinson and E. Vlieg {\it Surf. Science} {\bf 414} 159 (1998)
 \bibitem{Bailey} P. Bailey, T.~C.~Q. Noakes, D.~P. Woodruff, Surf. Sci. {\bf 426}, 358 (1999)
 \bibitem{Meier3} F. Meier, V. Petrov, H. Mirhosseini, L. Patthey, J. Henk, J. Osterwalder, and J. Hugo Dil, arXiv:1001.4927 (2010)
 \bibitem{Kaminski} D. Kaminski, P. Poodt, E. Aret, N. Radenovic, E. Vlieg, Surf. Sci. {\bf 575}, 233 (2005)
 \bibitem{Bentmann} H. Bentmann, F. Forster, G. Bihlmayer, E. V. Chulkov, L. Moreschini, M. Grioni and F. Reinert, Europhys. Lett. {\bf 87}, 37003 (2009)
 \bibitem{propto} This results from explicitly considering the lattice constant
 in the tight-binding calculation presented in Ref.
 \onlinecite{Petersen}.
 \bibitem{SbCuSpinSplitting} These values are estimated as an upper
 limit from the line width in the momentum distribution of the
 experimental band structure in Sb/Cu(111).
 \bibitem{Mirhosseini} H. Mirhosseini, J. Henk, A. Ernst, S. Ostanin, C.-T. Chiang, P. Yu, A. Winkelmann, and J. Kirschner, Phys. Rev. B {\bf 79}, 245428 (2009)
 \bibitem{Cercellier} H. Cercellier, C. Didiot, Y. Fagot-Revurat, B. Kierren, L. Moreau, and D. Malterre, and F. Reinert, Phys. Rev. B {\bf 73}, 195413 (2006)
 \bibitem{Cu} As we expect the spin-splitting in the Cu(111)
 surface state to be smaller than for the Ag(111) surface state,
 we have set the values to zero for the purpose of this work.
\end{thebibliography}
\end{document}